\documentclass[aps,prl,onecolumn,showpacs,showkeys,preprintnumbers]{revtex4}
\usepackage{eurosym}
\usepackage{amsmath}
\usepackage{dcolumn}
\usepackage{bm}
\usepackage{subfigure}
\usepackage{amsfonts}
\usepackage{amssymb}
\usepackage{makeidx}
\usepackage{epsfig}
\usepackage{graphicx}

\input{tcilatex}

\begin{document}

\title{\textbf{A new type of Multiverse, G\"{o}del theorems and the
nonstandard logic of classical, quantum mechanics and quantum gravity}}
\author{Massimo Tessarotto}
\affiliation{Department of Mathematics, Informatics and Geosciences, University of
Trieste, Via Valerio 12, 34127 Trieste, Italy}
\affiliation{Research Center for Theoretical Physics and Astrophysics, Institute of
Physics, Silesian University in Opava, Bezru\v{c}ovo n\'{a}m.13, CZ-74601
Opava, Czech Republic\\
Email: maxtextss@gmail.com}
\author{Claudio Asci}
\affiliation{Department of Mathematics, Informatics and Geosciences, University of
Trieste, Via Valerio 12, 34127 Trieste, Italy\\
Email: casci@units.it}
\author{Alessandro Soranzo}
\affiliation{Department of Mathematics, Informatics and Geosciences, University of
Trieste, Via Valerio 12, 34127 Trieste, Italy\\
Email: soranzo@units.it}
\author{Marco Tessarotto}
\affiliation{Central Directorate for Labor, Training, Education, and Family, Permanent
Structure for Innovation and Procedure Automation, Via San Francesco 37,
Trieste, Italy\\
Email: marcotts@gmail.com}
\author{Gino Tironi}
\affiliation{Department of Mathematics, Informatics and Geosciences, University of
Trieste, Via Valerio 12, 34127 Trieste, Italy\\
Email: gino.tironi@gmail.com}
\date{\today }

\begin{abstract}
The problem is posed of establishing a possible relationship between  new
types of \ Multi-verse representations, G\"{o}del undecidability theorems
and the logic of classical, quantum mechanics and quantum gravity. For this
purpose example cases of multi-verses are first discussed which arise in the
context of non-relativistic classical and quantum mechanics as well as
quantum gravity. As a result, it is confirmed that thanks to G\"{o}del
theorems non-relativistic classical and quantum mechanics, as well as
quantum gravity theory are incomplete. As a consequence, they necessarily
admit \ undecidable logical propositions and therefore obey a non-standard
three-way principle of non-contradiction, i.e., a propositional boolean
logic with the three different logical truth values. \emph{true}, \emph{false%
} and \emph{undecidable. }
\end{abstract}

\maketitle

\bigskip

\section{1 - Introduction}

A fundamental issue regarding the search non-standard logic in physics
concerns\ its possible consistent formulation, in the various appropriate
physical contexts, in terms of a \emph{unique non-standard boolean} \emph{%
logic, }namely characterized by discrete truth propositional values which
are consistent also with notion of undecidability due to G\"{o}del \cite%
{Goedel}. 

However, rather than investigating the possible role of G\"{o}del's
undecidability for a yet non-extant "theory of everything" \cite{kiefer}
(which should be able to encompass the whole physical phenomenology), here
we intend to look at existing theories of matter and space-time in order to
investigate their setting in the framework of boolean logic. \ More
precisely the aim of the present paper is to state\ the general validity of
the same \emph{non-standard \textbf{3-way }Principle of Non-Contradiction}
(non-standard 3-way PNC) which should hold for arbitrary physical systems
ranging from classical to quantum mechanics (CM and QM) as well for
classical and quantum gravity (CG and QG). 

The subject of the present paper is the \emph{possible connection of the
same non-standard PNC}, previously established for CM \cite{LOGIC-3}, QM 
\cite{LOGIC-1} and a suitable formulation of QG \cite{LOGIC-2}\ referred to
as covariant QG (CQG-theory \cite{cqg-3,cqg-4}), \emph{with the so-called} 
\emph{Undecidability Theorems} \emph{first formulated by G\"{o}del in 1931} 
\cite{Goedel}. 

This is a task which, perhaps surprisingly, was completely left unsolved by
Birkhoff and von Neumann (B-vN) in their famous 1936 paper appeared on
Annals of Physics \cite{Birkhoff-vonNeumann}. Nevertheless their paper,
dealing with the logic of non-relativistic QM and usually referred to in the
literature as the so-called "\emph{standard theory}" of quantum logic (QL),
was very influential for the subsequent development of mostly
mathematically-oriented research in quantum logic, dealing with mathematical
logic, model theory and topology \cite{Jammer,Corsia,Chang-Keisler}. This
has included also attempts to modify in various ways the B-vN formulation of
QL (see for example \cite{Ellermann2024} where it was suggested that QL
might be somehow replaced by the so-called "logic of partitions", namely
suitable equivalence relations among subsets \cite{Ellerrmann2010}).\ Such
an approach, however, just as the original contribution due to B-vN, is
incomplete because it is unrelated to the boolean logic (with the only
possible exception occurring in the so-called semiclassical limit). Finally,
in connection with the B-vN quantum logic, we should mention that another
possible route for the common logic of physics is that suggested/conjectured
in Ref.\cite{Morgan}. Although its actual feasibility remains to be seen,
this should imply that the logic of CM might be cast in a form similar to
the B-vN quantum logic and thus offer an alternative to the boolean logic
considered here.

On the other hand, as far as quantum logic is concerned, similar issues
arise also in the context of QG, a topic which was still uncharted at the
time of the original B-vN investigation. Therefore quantum logic should be
able to cope also with the complexity of QG, a subject for which several
different competing theoretical models exist and general criteria for their
possible selection are still debated \cite{Tessarotto2022}.

Our goal is to show that the connection with non-standard boolean logic
follows as a consequence of two \emph{Multi-verse representations} which can
be suitably associated with CM and respectively QM and QG.

In the literature several different types of Multi-verses are known \cite%
{Everett1957,
Albrecht,Carr,Czech,Johnson,Robles-Perez,Wallace,Weinberg,Garriga,Garriga
2006,Garriga 2009,Greene,Tegmark}. Particular relevance pertains here, for
its analogy, to the Everett Many Worlds (MW) representation of QM \cite%
{Everett1957}, which - nevertheless - has been regarded in the past either
incorrect \cite{Tegmark} or merely an \emph{abstract structure, }i.e., a
purely fictitious, unphysical. and non-observable feature, which cannot be
tested experimentally \cite{Ellis} and therefore should have no implication
whatsoever on the dynamics of CM, QM or QG.

According to the Everett's conjecture \cite{Everett1957}, QM should be
viewed as a manifestation of the existence of mutually non-interacting MW,
i.e., parallel universes, each one being regarded as purely classical and
thus obeying the classical evolution laws of CM. The trajectory of a single
quantum particle in these parallel universes, should therefore be uniquely
determined by prescribing as initial conditions the stochastic initial
values of a complete set of QM Lagrangian coordinates (which determine its
initial position). Given their classical character all such trajectories
should be considered equiprobable (see subsection 1.1). A variation along
the same lines is represented by the so-called Many Interacting Worlds
(MIW), a model due to Hall, Deckert and Wiseman \cite{EVERETT2-2} in which
the classical parallel universes are, instead, assumed as suitably mutually
interacting among them in terms of classical interactions of some sort.

In the present paper two different Multi-verse representations are
determined, respectively for the so-called Boltzmann-Sinai hard sphere
system \cite{CMFD-T,CMFD-T-1} (taken as a representative of CM),
non-relativistic QM (NRQM) and QG in the so-called covariant representation
provided by CQG-theory. These are here referred to as "\emph{Correlated Many
Worlds}" representations and, being related respectively to classical or
quantum systems, \ are therefore denoted respectively as \emph{Correlated
Classical Many Worlds} and \emph{Correlated Quantum Many Worlds }(\emph{CCMW}
and \emph{CQMW}). \ In analogy with the Everett's MW representation, they
are all based on the identification of appropriate \emph{generic
configuration domains, }i.e., abstract spaces which are spanned by suitable
stochastic Lagrangian coordinates\emph{.} However, in difference with
Everett's MW, the\ same Lagrangian coordinates are statistically correlated,
i.e., depend explicitly on suitable \emph{stochastic parameters}. The
prescription of the same parameters allows to determine \emph{specific }(or 
\emph{parallel}) \emph{configuration domains}, each one corresponding to a
prescribed value of the same stochastic parameters. It follows that each
stochastic Lagrangian coordinate, to be suitably prescribed, is uniquely
mapped onto a specific, or parallel, configuration space. \ 

In this reference we stress that, both for CM and QM, the choice of the
appropriate\ theoretical model appear of crucial importance in order to
identify both the generic and specific domains of the configuration-space
domain, i.e., the space spanned by the stochastic Lagrangian coordinates.
Thus, for example, in the case of non-relativistic QM (NRQM), only
theoretical models which are consistent with the so-called standard
interpretation of QM should be considered acceptable. This implies: 1) that
the customary Bohmian representation of QM must be ruled out because in such
a setting the Lagrangian trajectories are considered deterministic. 2) As a
consequence, in the Lagrangian representation of QM, the quantum Lagrangian
particle coordinates and corresponding trajectories (paths), when
represented in the generic configuration space (the Euclidean space $%
\mathbb{R}
^{3}$ which is spanned by all quantum point particles positions), should all
be considered as stochastic. This permits one to identify as a convenient
theoretical model for NRQM the so-called Generalized Lagrangian Path (GLP)
representation originally developed in Ref.\cite{Tessarotto2016} (see
discussion below in Section 2)..

Analogous considerations apply for QG. In this case a manifestly covariant
formulation of QG (referred shortly to as CQG theory \cite{cqg-3,cqg-4})
with respect to a background space-time representation can be adopted. As a
consequence, the theory of QG and the quantum Lagrangian variables -
together with the relevant stochastic parameters - can all be cast in
explicit $4-$tensor form with respect to the same background space-time. One
reason why a formulation of this type has been adopted for the formulation
of a boolean quantum logic for QG in Ref. \cite{LOGIC-2} is that in this
case the principle of general covariance is satisfied automatically, which
is also a prerequisite for quantum logic to make sense (the truth values
cannot depend on the choice of the coordinates). As far as the choice of the
Lagrangian variables is concerned, a further motivation is that, in analogy
to what happens in NRQM, also for QG a Bohmian representation of the
Lagrangian quantum trajectories should be ruled out because also in QG the
same trajectories are expected to be  non deterministic. A fact that allows
us to adopt a stochastic representation for the same Lagrangian coordinates.
As shown in Ref. \cite{Tessarotto 2018}, the  explicit $4-$tensor form of
CQG-theory allows it to be represented in terms stochastic Generalized
Lagrangian Paths, i.e., trajectories of the same variables (the so-called
GLP approach\ for CQG). It follows that also in this case the stochastic GLP
coordinates (and corresponding GLP trajectories) span an appropriate generic
configuration space. This is found to coincide with a ten dimensional
abstract space $%
\mathbb{R}
^{10},$ spanned by the variational symmetric tensor field $\mathbf{g}%
=\left\{ g_{\rho \nu }\right\} =\left\{ g^{\mu \nu }\right\} $(see
discussion below in Section 3).

As far as what concerns quantum logic, the identification of a possible
boolean realization requires of course the actual identification of discrete
logical truth values. As pointed out in Refs.\cite{LOGIC-1,LOGIC-2}, these
can be based on the quantum expectation values and related standard
deviations for quantum Lagrangian coordinates. Of course this means - as
suggested by the so-called \emph{Undecidability Theorems} established by a G%
\"{o}del \cite{Goedel} - that besides properly defined \emph{true }and \emph{%
false} logical truth values, also a further logical value, denoted as\emph{\
undecidable} (or\emph{\ uncertain}) should be suitably prescribed.

The task therefore involves the proper identification of the operative
physical definitions which are expected to generally apply only to specific
physical systems and to be realized in terms of suitable ideal\ (or \emph{%
Gedanken}) measurement experiments. The topic has been recently investigated
in various contexts, i.e., physical systems included in the following list
(not in chronological order):

\begin{itemize}
\item \emph{Boltzmann-Sinai system} (\emph{B-S system}) \cite{LOGIC-3}: an
example is realized by the Boltzmann-Sinai hard sphere system, which is
represented by an ensemble of $N>2$ hard-smooth spheres undergoing multiple
hard elastic collisions and in which the hard spheres are immersed in a
stationary and rigid parallelepiped domain $\Omega $,\ here denoted as \emph{%
generic configuration domain} (i.e., a bounded subset of the $3-$dimensional
Euclidean space $%
\mathbb{R}
^{3}$) to which the particle trajectories belong. Despite the fact that CL
is customarily associated with the CM of point particle systems, in Ref.\cite%
{LOGIC-3} we have shown that, on the contrary, CL does not generally apply
to the B-S system when the system hard-spheres undergo multiple collisions.
This explains why CL, which requires true propositions to be also unique,
becomes invalid for the Boltzmann-Sinai system in such a case and must be
replaced by a Boolean non-standard logic in which besides \emph{true }and%
\emph{\ false }logical propositions also \emph{undecidable} ones can occur.
The latter ones correspond to stochastic trajectories which, as shown in the
cited reference, generally occur (except in exceptional cases) after a
multiple collision event.

\item \emph{non-relativistic quantum mechanics }(\emph{NRQM}) \cite{LOGIC-1}%
: the example case is considered of a system of $N$ quantum point particles
with non-vanishing mass and immersed in the whole $3-$dimensional Euclidean
space $\Omega =%
\mathbb{R}
^{3},$ again identified with the \emph{generic configuration domain}$.$ For
definiteness let us consider the standard NRQM (i.e., in which quantization
is performed on the particle quantum positions $\mathbf{r}_{i}$), or
ontologically equivalent ones like the so-called GLP approach cited above 
\cite{Tessarotto2016} (which shortly is referred to here as GLP-NRQM). In
this case \emph{true }and \emph{false } logical propositions have been shown
to be respectively associated with a prescribed initial conditions (true),
or not (false), being represented by: A) the quantum expectation values of
the system particle positions $\mathbf{r}_{i}\equiv \left\{
r_{i}^{j},j=1,3\right\} $ of all particle, i.e., for all $i=1,N$ together
with B) finite values of the corresponding quantum standard deviations. In
addition, the \emph{undecidable} logical propositions are those for which
the corresponding standard deviation becomes infinite when evaluated in
terms of the prescribed initial conditions, i.e., the quantum measurement
becomes impossible.

\item \emph{covariant quantum gravity} (\emph{CQG}) \cite{LOGIC-2}: the
example is considered of a manifestly covariant formulation of QG (CQG
theory \cite{cqg-3,cqg-4}), endowed with a non-vanishing graviton mass and a
background metric tensor%
\begin{equation}
\widehat{\mathbf{g}}=\left\{ \widehat{g}_{\rho \nu }\right\} =\left\{ 
\widehat{g}^{\mu \nu }\right\}  \label{eqq-1}
\end{equation}%
which raises/lowers all tensor indices, and in which quantization is
performed w.r. to the variational metric tensor%
\begin{equation}
\mathbf{g}=\left\{ g_{\rho \nu }\right\} =\left\{ g^{\mu \nu }\right\} .
\label{eqq-2}
\end{equation}%
Thus $\mathbf{g}$ is defined so that its extremal value coincides with $%
\widehat{\mathbf{g}}$ while its\ tensor components (for example the
countervariant ones) are treated as unconstrained (i.e., are not required to
satisfy orthogonality conditions). Thus $\mathbf{g}$ is considered
effectively as a set of Lagrangian coordinates spanning a $k-$dimensional
real space $\Omega \equiv 
\mathbb{R}
^{10}$ to be identified again as \emph{generic configuration domain}$.$ Also
in this case a stochastic GLP approach can be formulated \cite{Tessarotto
2018}. The prescription of the \emph{true,} \emph{false }and \emph{%
undecidable} logical propositions is then formally analogous to the
treatment applying to NRQM, with $\mathbf{g}$ playing the role of the
configuration vector spanning the generic configuration space. More
precisely, the\ \emph{true/false} logical values are again associated with a
prescribed/unprescribed initial condition for the quantum wave-function,
being represented by: A) the quantum expectation values of the Lagrangian
coordinates $\mathbf{g}\equiv \left\{ g_{\rho \nu }\right\} $ of all
particle, i.e., for all $i=1,N$ together with B) finite values of the
corresponding quantum standard deviations. In addition, an \emph{undecidable}
logical proposition is the one in which the standard deviation becomes
infinite.
\end{itemize}

\bigskip

In the following we intend to investigate the Multi-verse features which
arise as a consequence of the stochastic nature of classical or quantum
particle trajectories. It should be noted that the first two cases share the
common feature that the particle trajectories belong to a
configuration-space which is identified either with the $3$D Euclidean space 
$%
\mathbb{R}
^{3}$ or a bounded subset of it. A side issue is to ascertain whether the
same Multi-verses are or are not of the Everett's MW type. According to the
Everett's conjecture \cite{Everett1957} this means that it should be
possible to represent all stochastic trajectories of arbitrary
(classical/quantum) systems in terms of the features indicated below.

\subsection{1.1 - FEATURES OF THE EVERETT MANY WORLDS REPRESENTATION}

\begin{enumerate}
\item (\emph{FEATURE \#1 MW: specific/parallel configuration space}) Each
specific/parallel configuration space is mapped by deterministic
trajectories. 

\item (\emph{FEATURE \#2 MW: equiprobability }) each parallel configuration
space should be considered independent of all the others. As a consequence
it follows that each trajectory and each parallel space should be considered
on an equal footing, i.e., equiprobable.

\item (\emph{FEATURE \#3 MW: classical deterministic trajectories}) the same
trajectories are also classical, i.e., determined purely by classical
interactions

\item (\emph{FEATURE \#4 MW: Hamilton equations}) each deterministic
trajectory can be described in terms of classical Hamilton equations.
\end{enumerate}

\section{2 - \protect\bigskip The Multi-verse representation of the S-B-S
system: CMW for S-B-S}

First, let us consider the case of the B-S hard sphere system in the case in
which multiple collisions are assumed to occur, namely the stochastic B-S
(S-B-S) system. We notice that the trajectories of the S-B-S system are by
construction stochastic on the configuration space the Euclidean space $%
\Omega \subset 
\mathbb{R}
^{3}$ (identified with a bounded parallelepiped with stationary and rigid
boundaries. This happens because of the following reasons:

\begin{itemize}
\item first, when multiple collisions occur (among $3$ or more hard
spheres), they are necessarily realized by suitable ordered sequences of
binary collisions (Prop.\#1 in \cite{LOGIC-2})\textbf{.}

\item second\textbf{,} multiple collisions can be realized by means of
different sequences of binary collisions (PROP. \#2. idem)

\item third, the same sequences generally lead to different "outgoing"(after
collision) states of the colliding particles (PROP. \#2 idem)

\item fourth, the exception is provided by "symmetric" multiple collisions
(PROP. \#3 idem)

\item fifth, except for case fourth, multiple collisions are characterized
by stochastic collision laws (PROP.\#4 idem)
\end{itemize}

As a consequence, after an arbitrary multiple collision event the $N$%
-particle hard sphere system is generally characterized by stochastic
trajectories, where each stochastic trajectory\ belongs to a different
specific/parallel configuration space. On the same parallel configuration
space the time evolution remains of course deterministic, until another
(possible) multiple collision even occurs. The number of stochastic
trajectories (and consequently of parallel configuration spaces) depends
therefore of the multiple-collision events which have occurred as well as on
the multiplicity ($f_{i}$) of each collision event (i.e., how many particles
are in each case simultaneously colliding). Thus, for example: in the case
of a double collision (i.e., in case of k=3 particles 1,2,3, in which
collisions (1-2) and (2-3) occur simultaneously) there are (k-1)!=2 possible
outgoing trajectories. Instead in the case of triple collisions ((i.e., in
case of k=4 particles 1,2,3,4 in which collisions (1-2),(2-3) and (3-4)
there are (k-1)!=6 different outgoing trajectories. Therefore the total
multiplicity ($f_{T}$) of stochastic trajectories at a given time $t$
depends on time-history, i.e., $f_{T}=f_{T}(t)$ . It follows therefore that
the ensemble of stochastic trajectories will generally be mapped on a \emph{%
Multi-verse} product space $\Omega ^{f_{T}(t)}.$

Finally, even if the different possible outgoing states occurring after a
given multiple collision event among $k$ particles are considered as endowed
with the same probability of occurrence $1/(k-1)!$ (i.e., equiprobable),
sequences of multiple collisions may lead to parallel configuration spaces
with different probability of occurrence. This means that the parallel
configuration spaces may be in a suitable sense "correlated", i.e. actually
realize an ensemble of correlated many worlds. Finally, the same worlds
should be considered also classical because particle dynamics after a
multiple collision event remains classical. Hence the label correlated
classical MW (CCMW).

Hence, the resulting Multi-verse abstract structure is realized by the \emph{%
correlated classical} MW (CCMW) representation of S-B-S (CCMW for S-B-S)
exhibits the features defined below.

\subsection{2.1 - FEATURES OF CORRELATED CLASSICAL MANY WORLDS REPRESENTATION%
}

The features of the CCMW Multi-verse representation are:

\begin{enumerate}
\item (\emph{FEATURE CCMW \#1: specific/parallel configuration space}) as an
ensemble of deterministic trajectories, with each one belonging to a "\emph{%
specific}" or "\emph{parallel}" configuration space\ and such that

\item (\emph{FEATURE CCMW \#2: correlated probability distribution, not
equiprobable}) the parallel configuration spaces (MW) may be statistically
correlated, i.e., have different probability of occurrence.

\item (\emph{FEATURE CCMW \#3: classical deterministic trajectories}) the
same trajectories are also classical, i.e., determined purely by classical
interactions, and therefore

\item (\emph{FEATURE CCMW \#4: classical Hamilton equations}) each
deterministic trajectory can be described in terms of classical Hamilton
equations.
\end{enumerate}

The comparison with the Everett MW representation is straightforward. In
fact while Features MW \#1, \#3 and \#4 are satisfied Feature MW \#2 is not
(generally) so because of the different possible statistical correlation and
the consequent (possible) violation of the equiprobability conditions.

\bigskip

\section{3 - \protect\bigskip The Multi-verse representation of GLP-NRQM:
CQMW for GLP-NRQM}

To begin with, we notice that in this context \cite{cqg-3,LOGIC-1} by
definition the quantum configuration space is identified with the Euclidean
space $\Omega =%
\mathbb{R}
^{3},$ on which the Lagrangian coordinates take value. For an $N-$body
(point particles) quantum system the latter are represented by the position
vector fields $\mathbf{r}_{i}$ (for $i=1,N$) of the $N$ quantum point
particles. On such a configuration space - as shown in Ref.\cite{LOGIC-2} -
the B-S system may exhibit stochastic, i.e., non unique trajectories which
span the same configuration

In fact based on the GLP representation cited above the quantum Lagrangian
trajectories (associated with the quantum point particles description
achieved in the context NRQM), trajectories spanning the Euclidean space $%
\Omega =%
\mathbb{R}
^{3}$, are by construction stochastic in terms of a constant, i.e.,
independent of $(\mathbf{r,}t),$ being defined up to a stochastic
displacement vector $\mathbf{\Delta r\equiv }\left( \Delta r_{1},\Delta
r_{2},\Delta r_{3}\right) ,$ which spans by construction the same space $%
\Omega =%
\mathbb{R}
^{3}$. Thus introducing the transformation%
\begin{equation}
\mathbf{r}_{i}\rightarrow \mathbf{R}_{i}=\mathbf{r}_{i}\ -\mathbf{\Delta r,}
\end{equation}%
The displaced position vectors $\mathbf{R}_{i}=\mathbf{r}_{i}$\ $-\mathbf{%
\Delta r}$, the corresponding trajectories $\mathbf{R}_{i}(t)$, denoted as
Generalized Lagrangian Paths (GLP) are therefore, by assumption, stochastic,
being characterized by a well-defined probability density. As a consequence
to each prescribed value of the stochastic vector $\mathbf{\Delta r}$ one
can associate a unique \emph{specific}/\emph{parallel} configuration space
on which the position (and corresponding deterministic trajectory) can be
mapped. Therefore a natural Multi-verse representation for NRQM follows once
the GLP representation (GLP-NRQM)\emph{\ }is adopted (the interested reader
may refer to Ref.\cite{Tessarotto2016} for background).

In fact now one can prove that again the GLP-NRQM theory admits a
Multi-verse representation. However, the notable difference is that both the
particle trajectories and Hamilton equations become quantum in nature.

This realizes a Multi-verse abstract structure which is achieved in terms of
a correlated \emph{quantum} MW (CQMW) representation of GLP-NRQM which
exhibits the features (distinct from those of the Everett's MW) indicated
below.

\subsection{3.1 - FEATURES OF CORRELATED QUANTUM MANY WORLDS REPRESENTATION}

The features of the CQMW Multi-verse representation

\begin{enumerate}
\item (\emph{FEATURE CQMW \#1: specific/parallel configuration space}) as an
ensemble of deterministic trajectories, with each one belonging to a "\emph{%
specific}" or "\emph{parallel}" configuration space\ and such that

\item (\emph{FEATURE CQMW \#2: correlated probability distribution, not
equiprobable}) the parallel configuration spaces (MW) may be statistically
correlated, i.e., have different probability of occurrence.

\item (\emph{FEATURE CQMW \#3: quantum trajectories}) the same trajectories
are also quantum, i.e., determined by quantum interactions, and therefore

\item (\emph{FEATURE CQMW \#4: quantum Hamilton equations}) each quantum
trajectory is described in terms of quantum Hamilton equations.
\end{enumerate}

\bigskip

\section{4 - \protect\bigskip The CMW-Multi-verse-verse representation of
CQG theory (CQMW for CQG theory)}

The covariant theory of QG (CQG theory \cite{cqg-3,cqg-4}) exhibits a strong
conceptual similarity with the GLP-NRQM theory. In fact also in this case
the configuration-space trajectories are, by construction, stochastic once a
stochastic GLP representation (GLP-CQG \cite{Tessarotto 2018}) is adopted
and taken of the form 
\begin{equation}
g_{\mu }^{\nu }(r)\rightarrow G_{\mu }^{\nu }(r)=g_{\mu }^{\nu }(r)-\Delta
g_{\mu }^{\nu }.
\end{equation}%
Here $\Delta \mathbf{g\equiv }$ $\left\{ \Delta g_{\mu }^{\nu }\right\} $%
(denoted again as \emph{displacement }$4-$\emph{tensor}) denotes a mixed
(covariant-countervariant) second order $4-$tensor which is assumed to be a
constant by construction (i.e., independent of $r\equiv \left\{
r^{i},i=0,3\right\} $) stochastic, symmetric and assumed to span the $10-$%
dimensional space $\Omega \equiv 
\mathbb{R}
^{10}$. In the framework of CQG, $G_{\mu }^{\nu }(r)$ and $g_{\mu }^{\nu
}(r) $ are regarded as second-order $4$-tensor whose indices are
raised/lowered by the background metric tensor $\widehat{\mathbf{g}}$ (see
Eq.(\ref{eqq-2})) which is parametrized in terms of the coordinates
(GR-frame) $r\equiv \left\{ r^{i},i=0,3\right\} $ \ Notice that here various
components $g_{\mu }^{\nu }(r)$ identify the Lagrangian coordinates, while $%
G_{\mu }^{\nu }(r)$ are the corresponding GLP variables$.$ In particular for
the GLP variables the set $\Omega $ coincides with the generic configuration
space spanned by all GLP-variables $G_{\mu }^{\nu }(r).$ \ Notice, in
addition, that in the framework of CQG, all the variational $4-$tensors
fields, i.e., $g_{\mu }^{\nu }(r)$ and $G_{\mu }^{\nu }(r),$ remain by
definition unconstrained and therefore are generally not metric tensors. As
a consequence the generic configuration space spanned by the stochastic
GLP-variables $G_{\mu }^{\nu }(r)$ coincides with the same $10-$dimensional
set $\Omega \subseteq 
\mathbb{R}
^{10}.$ Furthermore, to each prescribed value of the stochastic and constant 
$4-$tensor $\Delta g_{\mu }^{\nu }$ one can associate a unique \emph{specific%
}/\emph{parallel} configuration space on which the GLP-variables $G_{\mu
}^{\nu }(r)$ can be uniquely mapped. Therefore a natural Multi-verse
representation for CQG theory follows at once. \ A comparison with the
Everett's MW representation can be obtained also in this case. In fact one
can prove - based on the GLP-CQG approach developed in Ref.\cite{Tessarotto
2018} - the Multi-verse abstract structure is realized by the same CQMW
representation indicated above in subsection 3.1 (and hence trivially again
is not of the type of Everett MW).

\section{5 - Comparison with G\"{o}del Undecidability Theorems}

It is well known that G\"{o}del Undecidability Theorems \cite{Goedel} do not
hold for the logic of classical dynamical systems with configuration space
coinciding with the Euclidean configuration space $\Omega \subseteq 
\mathbb{R}
^{3}.$ An obvious example of this type is provided by Newtonian (Classical)
cosmology. Instead, different is the status of General Relativity in the
context of G\"{o}del Undecidability Theorems \cite{Barrow}.

The key feature of the three representations of Multi-verse obtained above\
(respectively for the S-B-S system, , and CQG) is, instead, that the
dimension of the corresponding CMW Multi-verses (respectively classical or
quantum) exceeds in all cases that of the Euclidean configuration space $%
\Omega \subseteq 
\mathbb{R}
^{3}$.

Indeed, in detail :

\begin{enumerate}
\item \emph{In case of the stochastic B-S system (it is assumed that at time 
}$t$\emph{\ multiple collisions have occurred in the same system)}: the CCMW
Multi-verse representation coincides with the product of the proper/parallel
configuration spaces $\Omega (\mathbf{r)}$, each one corresponding to a
prescribed position vector $\mathbf{r,}\prod\limits_{\alpha
=1,f_{T}(t)}\Omega (\mathbf{r)}\equiv \Omega ^{f_{T}(t)},$where $\ \Omega (%
\mathbf{r)\equiv }\Omega $ is a $3-$dimensional bounded subset of the
Euclidean space $%
\mathbb{R}
^{3}$ and $f_{T}(t)>1$ denotes the time-dependent total multiplicity ($f_{T}$%
) of stochastic trajectories at a given time $t$.

\item \emph{In the case the GLP representation of NRCM}: the CQMW
Multi-verse representation coincides with the infinite product of the
proper/parallel configuration spaces $\Omega (\Delta \mathbf{r)}$ each one
corresponding to a prescribed value of the stochastic vector $\Delta \mathbf{%
r,}$ $\Omega (\Delta \mathbf{r)}\equiv \Omega ,$with $\ \Omega $ the
Euclidean space $%
\mathbb{R}
^{3}$.

\item \emph{In the case of the GLP representation of CQG theory}: the CQMW
Multi-verse representation coincides with the infinite product of the
proper/parallel configuration spaces $\Omega (\Delta \mathbf{g)}$ each one
corresponding in this case to a prescribed value of the stochastic tensor $%
\Delta \mathbf{g,}$ $\Omega (\Delta \mathbf{g)}\equiv \Omega ,$with $\Omega $
denoting the space spanned by the variational symmetric tensor $\mathbf{g}$
and therefore coinciding with the $10-$dimensional Euclidean space $%
\mathbb{R}
^{10}.$
\end{enumerate}

One obtains therefore in each case, as a consequence, a sufficient condition
for the validity of G\"{o}del Undecidability Theorems \cite{Goedel}$.$ As a
result, the same theorems imply that all such systems are incomplete and
must admit undecidable logical sentences of some sort. Such an implication,
on the other hand, is therefore consistent with the common $3-$way PNC
previously established in Refs. \cite{LOGIC-1,LOGIC-2,LOGIC-3} for the same
systems (here listed in chronological order).

Therefore this confirms also:

\begin{itemize}
\item (that the 3-way PNC is not in contradiction with G\"{o}del's theory 
\cite{Goedel}, i.e., G\"{o}del Undecidability Theorems. In fact validity of
the $3-$way PNC is implied at once, even if the undecidable logical
propositions remain themselves arbitrary

\item the Multi-verse feature here pointed out and realized in terms of CMW
representations established above, is meaningful at least from the
conceptual\ viewpoint to establish the consistency with G\"{o}del's theory.

\item and finally, the \emph{universal character} of the $3-$way PNC,
holding both in classical and quantum physics.
\end{itemize}

Based on the fact that multiple collisions for this system may not generally
give rise to a unique outgoing state\ (i.e., defined after multiple
collisions), and therefore be deterministic in character, we intend to show
that multiple collision laws are generally stochastic, in the sense that
they may correspond to different - but equally probable - sequences of
simultaneous binary\ collisions events. This amounts to determine a
"statistical prescription" on the Boltzmann-Sinai hard-sphere system and to
analyze its basic qualitative properties, with special reference to the
determination of its logic.

\section{\protect\bigskip 6 - Conclusions}

The search of a boolean logics for quantum mechanics (QM) and quantum
gravity (QG) have been regarded for a long time fundamental as unsolved
problems of mathematical and theoretical physics. This includes also the
famous contribution due to Garrett Birkhoff and John von Neumann \cite%
{Birkhoff-vonNeumann} which is usually credited for establishing in the case
of NRQM the so-called \emph{standard }(approach to)\emph{\ quantum logic.}
However, despite its seminal achievements it did not actually solve the
problem,\ at least for the following two reasons:

\begin{itemize}
\item standard quantum logic is not pertain to boolean logic, and - in
particular -\ does not identify the appropriate notions of true and false
propositions to be based on the notions quantum expectation values and
standard deviations available in standard QM.

\item the possible identification of the undecidability (or indeterminacy)
logical truth value, relevant for establishing a possible realization of the
G\"{o}del incompleteness theorems was actually never achieved.
\end{itemize}

These are also most likely the reasons why, later on, von Neumann was
apparently unsatisfied by the outcome of their paper (quantum gravity was
still long time into the future). In fact, despite the fact that von Neumann
was well aware of the undecidability theorems established by\ G\"{o}del
already in 1931 (he personally attended the same International Conference of
Mathematics in K\"{o}nigsburg where G\"{o}del had first presented his famous
theorems), in his work with Birkhoff he missed any possible link with the
logic of propositional calculus based on Boolean logic.

In this paper such a connection has been established.

In fact the relationship of the Multi-verse representations holding
respectively in classical and quantum physics (i.e., quantum mechanics and
quantum gravity), with the boolean logic of classical mechanics,
non-relativistic QM and QG, together with G\"{o}del Undecidability Theorems
has been explained. \ For this purpose, first, Multi-verse representations
have been shown to hold for the stochastic Boltzmann-Sinai (the correlated
classical many worlds Multi-verse), and respectively for non relativistic
quantum mechanics and covariant quantum gravity\ (the correlated quantum
many worlds Multi-verse). To display the stochastic character of Lagrangian
coordinates (for NRQM and CQG) a stochastic representation, referred to as
generalized Lagrangian Path (GLP), has been adopted. \ 

The Multi-verse abstract representations obtained in this way are found in
all cases to differ from the customary Everett's Many Worlds (MW)
representation. In the case of the stochastic Boltzmann-Sinai this due to
the statistical correlations existing between different parallel
configuration domains. Instead for the considered quantum systems (NRQM and
CQG), the departure is also due to the quantum nature of quantum Hamilton
equation which in both cases includes the effect of Bohm quantum
interaction. As a consequences the configuration-space trajectories are
necessarily non-classical.

The results obtained here have important implications as far as the boolean
logic of classical mechanics, quantum mechanics and quantum gravity are
concerned. In fact, as a result, the validity and universal character of the
non standard, i.e., $3$-way, principle of non-contradiction (PNC),
established in such a framework, has been confirmed.

Finally, a brief comment regarding the broader perspective of logic in
general. The customary view is that the very concept of logic should be
rooted in the notion of classical logic in turn related to classical
cosmology, both relying on the $2-$way PNC due to Aristotle and set also at
the foundations of the Enlightenment philosophy \cite{Kant,ONLINE-IV-A} and
a broad range of modern philosophies. The present conclusions, indicate that
once the logic of classical and quantum physics is adopted, the logical
dichotomy true/false should be overcome. In this paper this point of view
has been put to its final challenge: the deterministic logic of classical
cosmology should be replaced by a boolean non-standard logic based on the $%
3- $way PNC.

\textbf{Data Availability Statement}

This manuscript has no associated data. [Authors' comment: The paper reports
theoretical analytical study. All data pertinent to this study are contained
in the paper.]

\textbf{Code Availability Statement}

This manuscript has no associated code/software. [Authors' comment: The
paper reports theoretical analytical study and no use of code/software is
made.]

\textbf{Declaration of competing interest}

The authors declare that to their knowledge there are no competing financial
interests or personal relationships that could have appeared to influence
the work reported in this paper.

\textbf{Funding information}

This research received no external funding.

\textbf{Conflict of Interest}

The authors declare no conflicts of interests.

\textbf{Author Contributions}

Conceptualization, Massimo Tessarotto, Claudio Asci, Alessandro Soranzo,
Marco Tessarotto and Gino Tironi; Investigation, Massimo Tessarotto, Claudio
Asci, Alessandro Soranzo, Marco Tessarotto and Gino Tironi; Writing --
original draft, Massimo Tessarotto, Claudio Asci, Alessandro Soranzo, Marco
Tessarotto and Gino Tironi. All authors have read and agreed to the
published version of the manuscript.

\end{document}